\documentclass{aastex701}
\usepackage{CJK, amsmath}

\begin{document}
\begin{CJK*}{UTF8}{gbsn}

\title{A Search for the Lost Comet P/2010 H2 (Vales)}

\correspondingauthor{Quanzhi Ye}

\author[orcid=0000-0002-4838-7676]{Quanzhi Ye (叶泉志)}
\affiliation{Department of Astronomy, University of Maryland, College Park, MD 20742, USA}
\affiliation{Center for Space Physics, Boston University, 725 Commonwealth Ave, Boston, MA 02215, USA}
\email[show]{qye@umd.edu}

\author[orcid=0000-0002-4767-9861]{Tony L. Farnham}
\affiliation{Department of Astronomy, University of Maryland, College Park, MD 20742, USA}
\email{farnham@astro.umd.edu}

\author{Perry Cai} 
\affiliation{Belmont High School, 221 Concord Ave, Belmont, MA 02478, USA}
\email{perrycai072@gmail.com}

\author[orcid=0000-0002-4230-6759]{Lori Feaga}
\affiliation{Department of Astronomy, University of Maryland, College Park, MD 20742, USA}
\email{feaga@astro.umd.edu}

\begin{abstract}

Short-period comet P/2010 H2 (Vales) underwent a significant outburst of $>7.5$~mag in 2010 and has not been detected since that apparition. Here we report our recovery attempt of P/Vales using the 4.3-m Lowell Discovery Telescope (LDT) during its 2015 and 2025 apparitions, as well as the data from the Transiting Exoplanet Survey Satellite (TESS) taken in 2023. With the LDT data, we did not detect the comet within the $3\sigma$ positional uncertainty ellipse to a $3\sigma$ limiting magnitude of $r\sim25$, corresponding to an absolute nuclear magnitude of $20.6$, or a diameter of $0.5$~km assuming a geometric albedo of 0.04. Similarly, the TESS data reveals no comet or debris trail, providing no direct evidence for a disruption event although not precluding one. The new constraint on the nucleus size tightens the range of viable activity mechanisms for P/Vales and is most consistent with a recently implanted, weakly processed nucleus. Our non-detection of P/Vales down to $m_r=25$ shows that objects like this are difficult to detect in their inactive state with Rubin Observatory, but shift-and-stack techniques and targeted observations on 10-m-class telescopes can provide more useful constraints on these objects.

\end{abstract}



\section{Introduction}

Outbursts in comets can signal large-scale changes on their nuclei, such as surface disruption or the exposure of fresh volatile material. While small outbursts within an amplitude of a magnitude or two are common among active comets, outbursts with amplitude exceeding a few magnitudes are comparatively rare, occurring less than once per year on average \citep{2016AJ....152..169I}. Such large outbursts indicate more powerful surface disruption events, providing unique opportunities to probe the compositional and structural properties of cometary nuclei.

Short-period comet P/2010 H2 (Vales) was discovered on UT 2010 April 16.0 as a stellar-appearing object at $V=12.6$ that apparently had brightened up from $V>20$ within 0.6 days \citep{2010ATel.2578....1B, 2010CBET.2249....1V}. The comet was tracked for five months until 2010 September, but was not seen again during subsequent apparitions in 2017 or 2025\footnote{See the data compiled by S. Yoshida, \url{http://www.aerith.net/comet/catalog/2010H2/index.html}, retrieved 2025 November 19.}. Analysis by \citet{2020PSJ.....1...77J} showed that the event released dust particles of $\sim10^9$~kg in mass, making it one of the most massive cometary outbursts since the outburst of 17P/Holmes in 2007. The same authors also placed an upper limit of $D=3$~km for the diameter of P/Vales.

As P/Vales has not been detected again, the aftermath of this spectacular outburst is unknown. In this paper, we report on our recovery attempt of P/Vales using the 4.3-m Lowell Discovery Telescope (LDT) at Happy Jack, AZ, as well as data from the Transiting Exoplanet Survey Satellite (TESS) satellite. Our observations aim to determine whether any remnant nucleus or dust activity remains detectable and, if absent, to constrain the current status of the remnant.

\section{Observations}

\begin{deluxetable*}{lccccccccl}
\tablecaption{Summary of observations used in this study.}
\label{tbl:obs}
\tablecolumns{7}
\tablehead{
\colhead{Date/Time (UT)} & \colhead{$r_\mathrm{h}$ (au)} & \colhead{$\varDelta$ (au)} & \colhead{$\alpha$} & \colhead{Telescope} & \colhead{Filter} & \colhead{Exposure} & \colhead{Image FWHM}
}

\startdata
2015 Dec 20 03:05--10:13 & 3.90 & 2.92 & $5.0^\circ$ & LDT & \textit{R} & $6\times300$~s & $1.7''$--$2.2''$ \\
2023 Oct 16 -- Nov 10 & 3.7--3.6 & 3.6--3.2 & $16^\circ$--$15^\circ$ & TESS & TESS & $10933\times200$~s & - \\
2025 Feb 28 10:03--11:27 & 3.08 & 2.31 & $13.6^\circ$ & LDT & \textit{VR} & $12\times180$~s & $2.1''$--$3.5''$ \\
2025 Apr 22 06:32--06:43 & 3.08 & 2.12 & $6.6^\circ$ & LDT & \textit{VR} & $3\times300$~s & $1.1''$--$1.2''$ \\
\enddata
\tablecomments{$r_\mathrm{h}$, $\varDelta$, and $\alpha$ are heliocentric distance, geocentric distance, and phase angle of the comet, respectively. FWHM = full-width-half-maximum.}
\end{deluxetable*}

\subsection{Lowell Discovery Telescope (LDT)}

We observed the ephemeris position of P/Vales computed using the latest JPL orbit solution on 2015 December 20, 2025 February 28, and 2025 April 22, using LDT's Large Monolithic Imager \citep[LMI;][]{Massey2021}. Observational circumstances are tabulated in Table~\ref{tbl:obs}. LMI has a field-of-view of $12\farcm3 \times 12\farcm3$ and an unbinned pixel scale of $0\farcs12$. The comet was observed with a Kron--Cousins $R$ filter for the 2015 December 20 observation, and an ultra-broadband {\it VR} filter for the remaining two dates. The 2015 run was divided into two sessions separated by 7 hours, and had total integration times of 15~minutes in each session, respectively; the two 2025 runs had total integration times of 1~hour and 15~minutes. Images were bias-subtracted, flat-field corrected, and were then astrometrically and photometrically calibrated using the Gaia DR3 catalog \citep{2023A&A...674A...1G} as well as the RefCat2 catalog \citep{2018ApJ...867..105T}, using the {\tt dct-redux} pipeline \citep{2024zndo..13946957K}. We used the Pan-STARRS $r$-band data to approximate our $R$- and {\it VR}-band observations.

To conduct a more robust search against contamination from background stars, we employed the image differencing technique to enhance faint object detection. For each night, we first stacked all frames from this night to create a ``reference'' image, and then subtracted each frame with this reference image using the {\tt HOTPANTS} algorithm \citep{2015ascl.soft04004B}. Finally, we combined all star-subtracted frames into one composite image following the motion of the comet.

\begin{figure}
    \centering
    \includegraphics[width=\linewidth]{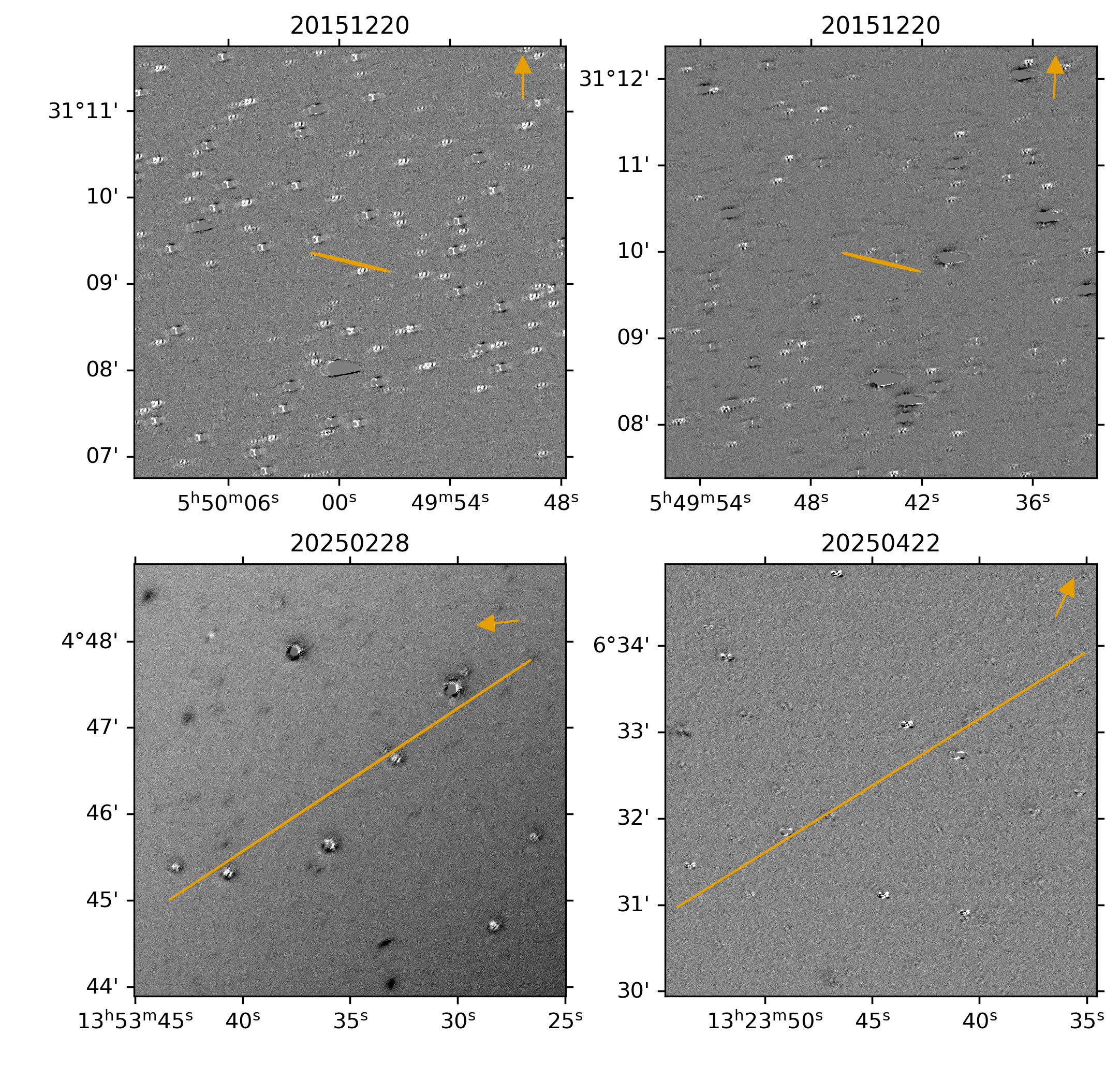}
    \caption{Star-subtracted composite of LDT images taken on 2015 December 20, 2025 February 28 and April 22. The highly elongated 3$\sigma$ uncertainty ellipses based on JPL orbit solution \#26 are plotted. Each image is $5'\times5'$ in size with celestial north at the top and the sunward direction denoted by the arrows.}
    \label{fig:ldt}
\end{figure}

The star-subtracted composite images are shown in Figure~\ref{fig:ldt}. The large frame-to-frame seeing variation, especially for the data taken on 2015 December 20 and 2025 February 28, likely contributed to imperfect kernel matching in the {\tt HOTPANTS} subtraction and produced significant residual artifacts. Despite so, our visual estimation shows that only $\lesssim3\%$ of the $3\sigma$ ellipse area is affected by these artifacts, and therefore they will not materially change our results. For each image, we searched the positional uncertainty ellipse of P/Vales computed using JPL's Orbit Solution \#26. We note that this orbit solution does not include parameters for non-gravitational accelerations primarily due to the short observational arc ($\sim5$~months). The $3\sigma$ limiting magnitudes for each date are calculated from the zero-points derived from the photometric calibration, which are found to be $m_r=24.03\pm0.05$, $24.6\pm0.1$, $25.26\pm0.03$, and $24.91\pm0.03$, for the two sessions on 2015 December 20, 2025 February 28, and 2025 April 22, respectively. Using $m_r=25.26$ (which provides the strictest constraint of the four sessions), the absolute nuclear magnitude of P/Vales can then be derived using the relation $M_2=m_r-5\log{\left(r_\mathrm{h} \varDelta \right)}-0.035\alpha=20.6$, in which $\alpha$ is the phase angle in degrees, and $0.035$ is the phase coefficient of cometary nuclei in magnitude per degree \citep{2004come.book..223L}. Assuming a geometric albedo of $0.04$ \citep{2004come.book..223L}, we derived upper limits on the nucleus diameter of $1.2$, $1.0$, $0.5$ and $0.5$~km for these four sessions.

\subsection{Transiting Exoplanet Survey Satellite (TESS)}

TESS is a telescope in Earth orbit that 
searches for transiting exoplanets by continuously observing a sector of the sky for 27 days \citep{2015JATIS...1a4003R}. With its wide field of view (four cameras, covering a $24^\circ\times96^\circ$ swath) TESS frequently observes comets that pass through its field of view.  The thousands of images obtained during the sector can be registered on the comet and then stacked to reveal faint coma \citep[][]{2021PSJ.....2..236F, 2023PSJ.....4...47Y}, dust trails \citep[][]{2019ApJ...886L..24F, 2025epsc.conf..775H}, and low surface brightness phenomena. P/Vales traversed TESS Sector~71 from 2023 October~16 - November 10  ($r_\mathrm{h}\sim3.6$~au, $\Delta\sim3.4$~au,  $\alpha\sim19^\circ$).  It appeared in a total of 10,933 usable 200-sec exposures (frames with extensive scattered light are discarded), resulting in a total integration time of 1.64$\times10^6$~s (cosmic ray filtering reduces the effective exposure time by 28\%). Following the approaches described in \citet{2019ApJ...886L..24F}, we removed the scattered light and background star field from individual images and then combined the data using a median stacking to minimize any residual star signals.  Data were coadded in various batches (1-day, 5-day and 25-day groups) to generate final images for further analysis.

\begin{figure}
    \centering
    \includegraphics[width=0.6\linewidth]{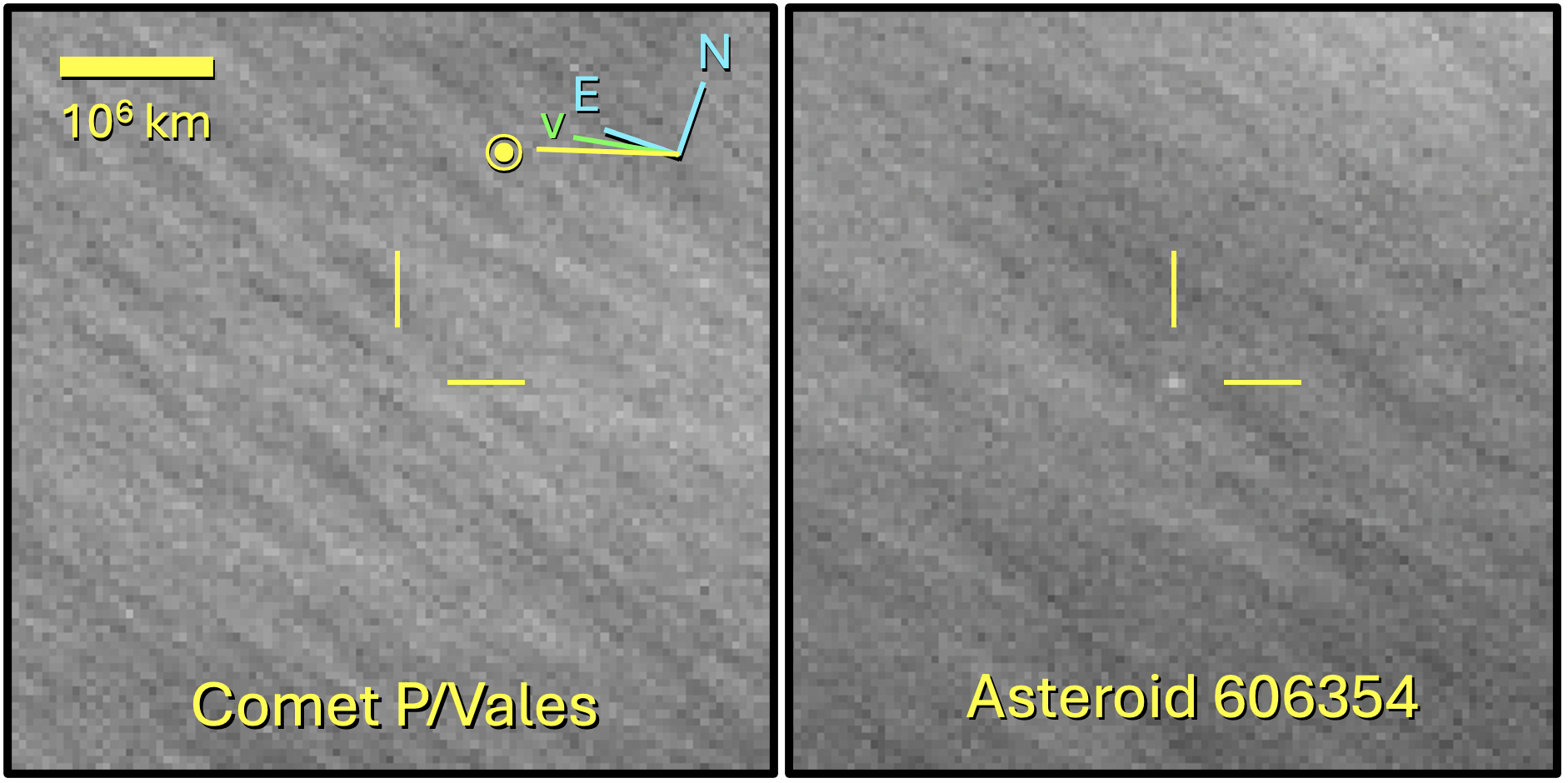}
    \caption{Coadded TESS frames illustrating the non-detection of comet P/Vales.  Left: Section of the stacked frame with all usable images registered at the comet's ephemeris position and then coadded.  The nominal ephemeris position of the nucleus is indicated, as well as the North and East directions, the sunward direction and the comet's velocity vector, $v$ (which also defines the orientation of the comet's orbit.  The $3\sigma$ uncertainty ellipse is 7 pixels in size and is not plotted in the interest of figure clarity.  Right: A comparison frame, with images registered on asteroid 606354, showing a clear detection of the $V$=21.9 object.}
    \label{fig:tess}
\end{figure}

Using the coadded TESS images, we searched for any sign of P/Vales.  Figure~\ref{fig:tess} (left panel) shows a subsection of the coadded frame produced from all of the usable images in the sector (with a total integration time of $1.64\times10^6$~s).  The predicted location of the comet is indicated, with a $3\sigma$ positional uncertainty of 7~pixels, but there is no sign of the object.  Given the characteristics of the TESS instruments, we should be able to obtain a clean detection of a body down to at least magnitude $V=22$ in our stacked image.  To test this limit, we performed our registration and coadding procedures on several nearby asteroids.  Figure~\ref{fig:tess} (right panel) shows one of these asteroids (606354), with a magnitude $V=21.9$, is clearly detected, suggesting that if P/Vales was at the nominal ephemeris position during the TESS observations, then it was fainter than $V=22$.

We performed additional analyses using the full field of view of the coadded TESS frames.  First, we searched for any object in the field that exhibited the same motion as the comet, in case it had drifted from the nominal ephemeris position since it was last observed (e.g., due to non-gravitational forces).  Second, we blinked multiple 5-day coadded frames to look for extended surface brightness structures (again offset from, but with the same motion as the comet) in case the nucleus had disrupted and left a residual debris field.  Both of these searches were negative.

The photometric limit from the TESS observations only restricts the nucleus diameter to $<$5~km, which is larger than the result from the LDT results.  However the TESS data do provide other valuable constraints on P/Vales.  The lack of a signal at the predicted position indicates that the comet did not exhibit significant activity around the time of the observations, and the lack of a detection anywhere in the field means that the comet is not simply offset from its expected position.  Finally, the absence of a debris field within $\sim10^8$~km of the nominal location of the comet means there is no direct evidence for a disruption event (though it does not preclude one).

\section{Discussion}

Our new constraint for the P/Vales nucleus diameter is more stringent than the constraint set by \citet{2020PSJ.....1...77J} using non-detections before the comet's 2010 outburst (0.5~km vs. 3~km), which also implies that the ejecta accounted for $\gtrsim4\%$ of the nucleus mass \citep[assuming the same bulk density of $500~\mathrm{kg~m^{-3}}$ used by][]{2020PSJ.....1...77J}, larger than previously estimated. Since both are only upper limits, it is inconclusive on whether P/Vales had experienced a significant size reduction or even a demise as a result of the outburst. However, the fact that P/Vales was tracked for five months after the outburst until it faded below the sensitivity limit of typical meter-class telescopes without reports of significant dispersion of ejecta, appears to suggest that the nucleus was still intact. Comets are known to survive powerful outbursts, sometimes without apparent fragmentation: 15P/Finlay experienced two strong outbursts during its 2014/15 apparition that ejected comparable scale of material to P/Vales's 2010 event \citep{2015ApJ...814...79Y, 2016AJ....152..169I}, 17P/Holmes survived its 2007 outburst which ejected $\sim3$~orders of magnitude more material \citep{2016ApJ...817...77I}, and 289P/Blanpain survived its 2013 outburst of comparable scale to P/Vales's 2010 event despite having one of the smallest nuclei ($D\sim0.3$~km) ever measured \citep{2019ApJ...878L..34Y}. We also note that if the 2010 outburst produced sustained asymmetric outgassing, the true long-term positional uncertainty could exceed the formal $3\sigma$ region derived from a purely gravitational solution as in the JPL orbit solution. However, the lack of significant residual trends in the 2010 astrometry and the rapid fading of the coma argue against prolonged strong non-gravitational forcing.

\citet{2020PSJ.....1...77J} suggested that the outburst was most likely triggered by conductive heating of subsurface ice, possibly through exothermic crystallization. This is supported by the detection of absorption features within the ejecta related to crystalline ice \citep{2010DPS....42.0509Y} as well as the high speeds of the ejecta \citep{2020PSJ.....1...77J}. To expel $M_\mathrm{d}=10^9$~kg of dust \citep{2020PSJ.....1...77J}, it requires an energy of $\frac{1}{2}M_\mathrm{d}v^2\approx1\times10^{12}~\mathrm{J}$, taking $\bar{v}=50~\mathrm{m/s}$ appropriate for a characteristic dust size of $\bar{a}=10~\micron$ using the speed derived by coma modeling \citep{2020PSJ.....1...77J}. Using the crystallization energy release for amorphous ice of $\sim10^5~\mathrm{J/kg}$ \citep[cf.][]{2024come.book..823P}, it requires $\sim10^7$~kg of crystallized ice to expel $\sim10^9$~kg of dust at high speed, corresponding to a shell of a few meters thick on average on a $D=0.5$~km nucleus. While this remains plausible, it becomes increasingly restrictive for a smaller nucleus: if the outburst was powered by the crystallization of a substantial reservoir of amorphous ice, then P/Vales must have experienced limited prior thermal processing, as small nuclei have short thermal conduction and rotational excitation timescales and tend to disrupt rapidly \citep{1997EM&P...79...35J, 2004come.book..223L}. The retention of meter-scale amorphous ice layers on such a body would therefore favor a relatively pristine object that has only recently reached its current orbit. Despite being in the outer main asteroid belt, P/Vales is dynamically distinct from the nearby Hilda asteroids that are clustered around the 3:2 mean-motion resonance with Jupiter, supporting the idea that the comet is dynamically unstable and likely only recently arrived the inner Solar System.

The non-detection as well as the apparent lack of renewed activity exhibited by P/Vales near perihelion after 2010 is not necessarily surprising, given the perihelion distance of $q > 3$ au. At such heliocentric distances, water ice sublimation rates are low, and activity may instead be driven by more volatile species (e.g., CO or CO$_2$) or by episodic crystallization events. If the 2010 outburst was triggered by crystallization of a localized reservoir of amorphous ice, subsequent activity would depend on whether additional reservoirs remain accessible. A small nucleus ($\sim0.5$~km) could have exhausted a near-surface volatile pocket during the 2010 event, leading to quiescence in later returns.

At $D\sim0.5$~km, an inactive P/Vales would only reach $V\sim25$ even during favorable oppositions, requiring either long integrations on 10-m-class ground-based telescopes or smaller but space-based telescopes for detection. Such faint objects would be below the single-exposure sensitivity of the Vera C. Rubin Observatory or NEO Surveyor, whose survey strategy and data processing are optimized for identifying moving objects in single exposures \citep{2023ApJS..266...22S, 2023PSJ.....4..224M}. Finding/recovering inactive objects/remnants of this size will therefore require alternative approaches, such as shift-and-stack techniques or dedicated search strategies. Future searches that combine Rubin's or NEO Surveyor's survey depth and cadence with targeted deep imaging from existing or new wide-field instruments on 10-m-class telescopes would be critical for studying objects like P/Vales and understanding how common such objects are.

\begin{acknowledgments}
We thank two anonymous referees for their careful reading and comments, as well as Hayden Green, Beck Maier, Ishara Nisley, Teznie Pugh, and Cecilia Siqueiros for assisting the Lowell Discovery Telescope (LDT) observations. The LDT observations were acquired by the University of Maryland observing team consisting of Gerbs Bauer, Adeline Gicquel-Brodtke, Tony Farnham, Lori Feaga, Michael Kelley, Jessica Sunshine, and Quanzhi Ye. These results made use of LDT at Lowell Observatory. Lowell is a private, non-profit institution dedicated to astrophysical research and public appreciation of astronomy and operates the LDT in partnership with Boston University, the University of Maryland, the University of Toledo, Northern Arizona University and Yale University.
\end{acknowledgments}

%
\facilities{LDT, TESS}

\software{{\tt dct-redux} \citep{2024zndo..13946957K}, {\tt HOTPANTS} \citep{2015ascl.soft04004B}}



\bibliography{sample701}{}
\bibliographystyle{aasjournalv7}



\end{CJK*}
\end{document}